\documentclass[prd,aps,showpacs,epsf,floats,onecolumn]{revtex4}%
\usepackage{amssymb}
\usepackage{amsfonts}
\usepackage{amsmath}
\usepackage{graphicx}%
\providecommand{\U}[1]{\protect\rule{.1in}{.1in}}
\begin{document}
\title{\textbf{Upper Limit on the Acceleration of a Quantum Evolution in Projective
Hilbert Space}}
\author{\textbf{Paul M. Alsing}$^{1}$ and \textbf{Carlo Cafaro}$^{2,3}$}
\affiliation{$^{1}$Air Force Research Laboratory, Information Directorate, Rome, NY 13441, USA}
\affiliation{$^{2}$University at Albany-SUNY, Albany, NY 12222, USA}
\affiliation{$^{3}$SUNY Polytechnic Institute, Utica, NY 13502, USA}

\begin{abstract}
It is remarkable that Heisenberg's position-momentum uncertainty relation
leads to the existence of a maximal acceleration for a physical particle in
the context of a geometric reformulation of quantum mechanics. It is also
known that the maximal acceleration of a quantum particle is related to the
magnitude of the speed of transportation in projective Hilbert space. In this
paper, inspired by the study of geometric aspects of quantum evolution by
means of the notions of curvature and torsion, we derive an upper bound for
the rate of change of the speed of transportation in an arbitrary
finite-dimensional projective Hilbert space. The evolution of the physical
system being in a pure quantum state is assumed to be governed by an arbitrary
time-varying Hermitian Hamiltonian operator. Our derivation, in analogy to the
inequalities obtained by L. D. Landau in the theory of fluctuations by means
of general commutation relations of quantum-mechanical origin, relies upon a
generalization of Heisenberg's uncertainty relation. We show that the
acceleration squared of a quantum evolution in projective space is upper
bounded by the variance of the temporal rate of change of the Hamiltonian
operator. Moreover, focusing for illustrative purposes on the
lower-dimensional case of a single spin qubit immersed in an arbitrarily
time-varying magnetic field, we discuss the optimal geometric configuration of
the magnetic field that yields maximal acceleration along with vanishing
curvature and unit geodesic efficiency in projective Hilbert space. Finally,
we comment on the consequences that our upper bound imposes on the limit at
which one can perform fast manipulations of quantum systems to mitigate
dissipative effects and/or obtain a target state in a shorter time.

\end{abstract}

\maketitle

\section{Introduction}

A fundamental task in quantum physics is that of finding the maximal speed of
evolution of a quantum system. In general, the maximal speed is obtained by
reducing the time or speeding up the quantum process. For a review on the
topic, we refer to Refs. \cite{frey16,deffner17}. These so-called quantum
speed limits (QSLs) were originally presented for closed systems evolving
unitarily between orthogonal states by Mandelstam-Tamm (MT) in Ref.
\cite{mandelstam45} and by Margolus-Levitin (ML) in Ref. \cite{margo98}. These
two bounds are given by,%
\begin{equation}
\tau_{\mathrm{QSL}}=\max\left\{  \pi\hslash/(2\Delta E)\text{, }\pi
\hslash/\left[  2(E-E_{0})\right]  \right\}  \text{,} \label{bound}%
\end{equation}
respectively. The MT bound follows from the Heisenberg uncertainty relation
and depends on the variance $\Delta E$ of the energy of the initial state. The
ML bound relies on the concept of the transition probability amplitude between
two quantum states in Hilbert space and depends on the mean energy $E$ of the
initial state with respect to the ground state energy $E_{0}$. Observe that
for a fixed average energy $E$, $\Delta E$ can be arbitrarily large. For a
discussion on extensions of QSL bounds to time-dependent Hamiltonian
evolutions between two arbitrary orthogonal states and to nonunitary
evolutions of open quantum systems, we refer to Refs. \cite{frey16,deffner17}.
For a discussion on extensions of the MT quantum speed limit to systems in
mixed states, we suggest \cite{ole22}. Since when $\hslash$ goes to zero, the
MT and ML quantum speed limit times approach zero as evident from Eq.
(\ref{bound}), one might be inclined to think that speed limits are a purely
quantum effect and no \textquotedblleft classical\textquotedblright\ speed
limit exists. Remarkably, it can be shown that QSL as a universal property
that occurs in both quantum and classical domains. Indeed, speed limits are
not derived from quantum noncommutativity. Instead, they emerge as a
consequence of peculiar dynamical properties of Hilbert space. Classical speed
limits, in particular, emerge once one uses a phase-space formulation of
quantum mechanics \cite{delcampo18} or, alternatively, a Hilbert space
formulation for the classical Liouville equation \cite{ohzeki18}. In
classical, semiclassical, and quantum settings, speed limits are specified by
a given norm of the generator of the dynamics and the state of the system
being analyzed.

In addition to speed limits, higher-order rates of changes can be equally
relevant in quantum physics. In this respect, the geometric mechanics works by
Caianiello and collaborators in Refs. \cite{cai81,cai82,cai84} on the maximal
acceleration of a quantum particle are of special significance. Furthermore,
following the investigations of Caianiello and collaborators in Refs.
\cite{cai81,cai82,cai84}, the works by Pati in Refs. \cite{pati92,pati92b}
deserve to be mentioned since they demonstrate the rather intriguing result
according to which the maximal acceleration of a quantum particle is related
to the magnitude of the velocity of transportation in projective Hilbert
space. Going beyond the concept of acceleration, one has the notion of jerk,
the time-derivative of the acceleration \cite{schot78,eager16}. The jerk
appears in classical electrodynamics in the so-called Abraham-Lorentz equation
describing the radiative loss of energy of an accelerated charged particle
\cite{murphy77,rohrlich00}. In cosmology, the concept of \textquotedblleft
cosmic jerk\textquotedblright\ is referred to the discovery of the
accelerating expansion of the Universe through observations of distant
supernovae \cite{riess98,riess04}. For illustrative examples of the notion of
jerk for curves in three-dimensional Euclidean space (emerging, for instance,
from the motion of an electron in the presence of a constant magnetic field)
equipped with moving Frenet-Serret frames, we refer to Refs.
\cite{ozen19,ozen20}. As a predictor of large accelerations of short
durations, the jerk has important practical applications in engineering,
especially in robotics and automation \cite{saridis88}. In these classical
engineering applications, one tends to minimize the jerk. This minimization is
motivated by the fact that the efficiency of control algorithms is negatively
affected by the third derivative of the position along the desired trajectory
\cite{saridis88}. Recently, minimum jerk trajectory techniques \cite{wise05}
have been employed in quantum settings to obtain optimal paths along which
efficiently transport atoms in optical lattices \cite{liu19,matt23}. Indeed,
finding efficient ways to control and manipulate atoms with minimal losses
represent essential steps toward the coherent formation of single ultracold
molecules suitable for explorations in quantum information processing and
quantum engineering.

In this paper, motivated (at the origin) by the existence of an upper bound on
the acceleration of a quantum particles as shown by Caianiello and
collaborators in Refs \cite{cai81,cai82,cai84}, we discuss two main results:

\begin{enumerate}
\item[{[i]}] An upper bound on the acceleration of transportation in an
arbitrary finite-dimensional projective Hilbert space in terms of the
square-root of the variance of the time-derivative of the Hermitian
Hamiltonian operator that governs the evolution of the system.

\item[{[ii]}] An upper bound on the time-derivative of the maximal
acceleration of a quantum particle (i.e., its maximal jerk) in terms of the
acceleration of transportation in projective Hilbert space and, thus (because
of [i]), in terms of the standard deviation of the first-derivative of the
Hermitian Hamiltonian operator of the system itself.
\end{enumerate}

The result in [i] is essentially a consequence of a generalization of
Heisenberg's uncertainty relation which, in turns, leads to the important
inequalities obtained by L. D. Landau in the theory of fluctuations \ and
exploited by Pati in Refs. \cite{pati92,pati92b}. The result in [ii], instead,
is a consequence of the result in [i] and of the link between acceleration of
a quantum particle and speed of transportation in projective Hilbert space as
reported by Pati in Ref. \cite{pati92}.

The layout of the rest of this paper is as follows. In Section II, we
introduce the concept of maximal acceleration for a quantum particle. First,
we begin with Caianiello's 1981 proposal based on a eight-dimensional
relativistic phase-space description of a quantum particle \cite{cai81,cai82}%
.\ Second, we revisit Caianiello's 1984 proposal based on the Heisenberg
time-energy uncertainty relation \cite{cai84}. In Section III, we discuss the
solution to the criticisms moved against Caianiello's 1984 proposal as
proposed by Pati in 1992 \cite{pati92}. In view of their usefulness in our own
derivations in Section IV, we not only consider Pati's derivation of the
maximal acceleration in Section II, but we also revisit important inequalities
within Landau's theory of fluctuations along with the link between maximal
acceleration and maximal energy loss for a quantum particle \cite{pati92b}. We
conclude Section III with a revisitation of Pati's proposed link between the
maximal acceleration of a quantum particle and the speed of transportation in
projective Hilbert space. After the critical revisitation of the material
presented in Sections II and III, we finally provide in Section IV our novel
upper bound on the acceleration in projective Hilbert space in terms of the
square-root of the dispersion (or, alternatively, the variance) of the
time-derivative of the nonstationary Hamiltonian of the finite-dimensional
quantum system being studied. We conclude Section IV with a practical
discussion on the consequences of our upper bound for the maximal jerk of a
spin-$1/2$ particle as a single-qubit quantum system. Our conclusive remarks
appear in Section V. Finally, some technical details are placed in Appendix A.

\section{Maximal acceleration of a quantum particle}

In this section, we introduce here the concept of maximal acceleration for a
quantum particle. We begin by revisiting Caianiello's 1981 proposal based on a
eight-dimensional relativistic phase-space description of a quantum particle
\cite{cai81,cai82}.\ We then present Caianiello's 1984 proposal based on the
Heisenberg time-energy uncertainty relation \cite{cai84}. In particular, we
discuss the criticisms moved against this latter proposal.

\subsection{Caianiello's 1981 proposal}

It is known that the positivity of the line element specified by the
Minkowskian metrization of the four-dimensional space-time,%
\begin{equation}
ds_{\mathrm{Minkowski}}^{2}\overset{\text{def}}{=}c^{2}dt^{2}-d\vec{x}%
^{2}\text{,} \label{Min}%
\end{equation}
implies that no physically meaningful speed can be greater than the speed of
light $c$. Interestingly, the classical theory of special relativity does not
impose any limit on the acceleration of a particle. In Refs.
\cite{cai81,cai82}, Caianiello proposed a physically motivated geometric
description of a quantum particle in a eight-dimensional relativistic
phase-space extended with both time $t$ and energy $E$ dimensions, along with
a line element given by%
\begin{equation}
ds_{\mathrm{Caianiello}}^{2}\overset{\text{def}}{=}c^{2}dt^{2}-d\vec{x}%
^{2}+\frac{\hslash^{2}}{\mu^{4}c^{4}}\left[  \frac{1}{c^{2}}dE^{2}-d\vec
{p}^{2}\right]  \text{.} \label{Cai}%
\end{equation}
In Eq. (\ref{Cai}) $\hslash\overset{\text{def}}{=}h/(2\pi)$ and $\mu$ is a
mass such that $\sigma_{p}\overset{\text{def}}{=}\mu c$ specifies the minimum
momentum uncertainty, with $\sigma_{x}\sigma_{p}=\hslash$. Moreover,
$\sigma_{x}\overset{\text{def}}{=}\lambda$ is the minimum position uncertainty
and, finally, $\lambda$ denotes the linear dimension of the quantum particle
viewed as an extended object. For completeness, we recall that in the context
of Born's reciprocity theory of elementary particles \cite{born49}, the
quantities $\lambda$ and $\mu c$ denote a fundamental \textquotedblleft
length\textquotedblright\ and \textquotedblleft momentum\textquotedblright,
respectively, used to split the reduced Planck constant as\textbf{ }%
$\hslash=\left(  \lambda\right)  \left(  \mu c\right)  $\textbf{.} Recall that
the position and momentum four-vectors in relativity are given by $x^{\beta
}\overset{\text{def}}{=}\left(  ct\text{, }\vec{x}\right)  $ and $P^{\beta
}\overset{\text{def}}{=}\left(  E/c\text{, }\vec{p}\right)  $, respectively,
with $0\leq\beta\leq3$. The energy $E$ and the three-vector $\vec{p}$ are
given by $mc^{2}\overset{\text{def}}{=}c\sqrt{m_{0}^{2}c^{2}+\vec{p}\cdot
\vec{p}}$ and $m\vec{v}\overset{\text{def}}{=}\gamma\left(  \vec{v}\right)
m_{0}\vec{v}$, respectively. The quantity $m_{0}$ denotes the rest mass of the
particle, $\vec{v}\overset{\text{def}}{=}d\vec{x}/dt$, $m\overset{\text{def}%
}{=}\gamma\left(  \vec{v}\right)  m_{0}$, and $\gamma\left(  \vec{v}\right)
\overset{\text{def}}{=}\left[  1-(\vec{v}\cdot\vec{v})/c^{2}\right]  ^{-1/2}$
is the Lorentz factor in relativity. In Caianiello's geometric proposal, both
the velocity and the acceleration four-vectors of the particle play a
meaningful role since there is the interest in measuring both $x^{\beta}$ and
$P^{\beta}$. Indeed, from the positivity of the line element in Eq.
(\ref{Cai}), Caianiello obtains%
\begin{equation}
a\leq\frac{\mu}{m_{0}}\frac{c^{2}}{\lambda}\text{,} \label{cai1}%
\end{equation}
where $a\overset{\text{def}}{=}\left\Vert \vec{a}\right\Vert $ and $\vec
{a}\overset{\text{def}}{=}d\vec{v}/dt$. The significance of Eq. (\ref{cai1})
is encoded in the fact that the extended nature of a quantum particle limits
its acceleration in a fundamental way. If the particle is point-like, its
linear dimension $\lambda$ would be zero and its acceleration would have no
upper bound. Interestingly, the extended nature of particles is not unusual in
the literature \cite{caticha}. In particular, it was also proposed in the
context of a relativistic quantum particle in Ref. \cite{hegerfeldt74} where
it was shown that the localization of a particle at a given space-time point
violates causality. As a final remark, we point out that the existence of an
upper limit on the acceleration of massive particles in the framework of
quantum geometry proposed by Caianiello was exploited in Ref.
\cite{capozziello00} to derive the generalized uncertainty principle of string theory.

\subsection{Caianiello's 1984 proposal}

In a subsequent work appeared in Ref. \cite{cai84}, Caianiello tackled an
alternative derivation of Eq. (\ref{cai1}) and claimed that the existence of a
maximal acceleration for physical particles was a consequence of Heisenberg's
time-energy uncertainty relation. He concluded that%
\begin{equation}
a\overset{\text{def}}{=}\left\vert \frac{dv}{dt}\right\vert \leq\frac
{2}{\hslash}\left(  \Delta E\right)  \left(  \Delta v\right)  \text{,}
\label{cai2}%
\end{equation}
where $\Delta v\overset{\text{def}}{=}\sqrt{\left\langle \left(
v-\left\langle v\right\rangle \right)  ^{2}\right\rangle }=\sqrt{\left\langle
v^{2}\right\rangle -\left\langle v\right\rangle ^{2}}$ is the uncertainty in
the velocity of the particle with energy $E$ and energy uncertainty $\Delta
E$. In particular, assuming $\Delta E\leq E$ with $E\simeq m_{0}c^{2}$ for a
particle nearly at rest and noting that $\Delta v\leq v_{\mathrm{\max}}\leq
c$, Caianiello concluded that%
\begin{equation}
a_{\mathrm{\max}}=2\frac{m_{0}c^{3}}{\hslash}\text{.} \label{cai3}%
\end{equation}
Caianiello's derivation of Eqs. (\ref{cai2}) and (\ref{cai3}) was criticized
for three reasons. First, Eq. (\ref{cai2}) happens to be correct, but it is
not a consequence of the time-energy uncertainty principle. Instead, as shown
in Ref. \cite{wood89}, it is a consequence of an inequality obtained in the
context of Landau's theory of fluctuations \cite{landau1,landau2}. This
inequality, in turn, can be derived by using standard quantum-mechanical
methods along with the \emph{ansatz} between the quantum commutator and the
classical Poisson bracket, $\left[  \cdot\text{, }\cdot\right]  /(i\hslash
)\rightarrow\left\{  \cdot\text{, }\cdot\right\}  $. Second, as pointed out by
Papini in Ref. \cite{papini03}, the assumption $\Delta E\leq E$ is generally
false. Indeed, given a fixed average energy $E$, a quantum state can have an
arbitrarily large $\Delta E$ as evident from the time-energy uncertainty
relation $\Delta E\geq\hslash/(2\Delta t)$. Third, as pointed out by Sharma in
Ref. \cite{sharma85}, the rest mass $m_{0}$ in Eq. (\ref{cai3}) should be
generally replaced by $m=\gamma\left(  \vec{v}\right)  m_{0}$. However, $m$ is
not invariant under flat Lorentz transformations. In addition, even replacing
$m_{0}c^{2}$ with $mc^{2}$, one can notice that when $v$ approaches $c$, $E$
is not bounded and, therefore, the acceleration of the particle has no longer
an upper limit. In the next section, we revisit the manner in which Pati
addressed the criticisms moved against Cainiello's 1984 proposal.

\section{Pati's 1992 derivation of Cainiello's 1984 proposal}

In this section, we discuss a solution to the criticisms moved against
Caianiello's 1984 proposal as suggested by Pati in 1992 \cite{pati92}. First,
we review an important inequality within Landau's theory of fluctuations. This
inequality is essential in Pati's rederivation of Caianiello's 1984 proposal.
Second, while going over Pati's rederivation, we emphasize the link between
maximal acceleration and maximal energy loss for a quantum particle
\cite{pati92b}. Finally, we reconsider Pati's suggested link between the
maximal acceleration of a quantum particle and the speed of transportation in
projective Hilbert space. The inequality within Landau's theory of
fluctuations along with this link between the maximal acceleration of a
quantum particle and the speed of transportation in projective Hilbert space
are essential to our own work that will appear in Section IV.

\subsection{The inequality in Landau's theory of fluctuations}

As previously mentioned, this inequality is essential in Pati's rederivation
of Caianiello's 1984 proposal. Following Ref. \cite{pati92b}, consider two
observable operators $\hat{f}$ and $\hat{g}$ with variances defined with
respect to the state $\left\vert \Phi\right\rangle $ given by%
\begin{equation}
\left(  \Delta f\right)  ^{2}\overset{\text{def}}{=}\left\langle
\Phi\left\vert \left(  \hat{f}-\left\langle \hat{f}\right\rangle \right)
^{2}\right\vert \Phi\right\rangle \text{, and }\left(  \Delta g\right)
^{2}\overset{\text{def}}{=}\left\langle \Phi\left\vert \left(  \hat
{g}-\left\langle \hat{g}\right\rangle \right)  ^{2}\right\vert \Phi
\right\rangle \text{,}%
\end{equation}
respectively. From quantum mechanics \cite{sakurai85}, we have that%
\begin{equation}
\left(  \Delta f\right)  \left(  \Delta g\right)  \geq\frac{1}{2}\left\vert
\left\langle \Phi\left\vert \left[  \hat{f}\text{, }\hat{g}\right]
\right\vert \Phi\right\rangle \right\vert \text{.} \label{impo}%
\end{equation}
For a more in depth comment on the difference between the pairs $\left(
f\text{, }g\right)  $ and $($ $\hat{f}$, $\hat{g})$, we refer to our
observations in \cite{comment}. In the spirit of recovering an uncertainty
relation in the quasi-classical limit \cite{landau1,landau2}, let us put
$\left[  \hat{f}\text{, }\hat{g}\right]  =-i\hslash\hat{c}$ with $\hat{c}$ a
Hermitian operator. Recall the ansatz between the quantum commutator and the
classical Poisson bracket, $\left[  \hat{f}\text{, }\hat{g}\right]
/(i\hslash)\rightarrow\left\{  f\text{, }g\right\}  $. Then, setting $\hat
{c}=\left\{  f\text{, }g\right\}  \mathbf{1}$ with $\mathbf{1}$ being the
identity operator, assuming $\hat{f}=\mathrm{\hat{H}}$ and, finally, using
ansatz along with Eq. (\ref{impo}), we get%
\begin{equation}
\left(  \Delta E\right)  \left(  \Delta g\right)  \geq\frac{\hslash}%
{2}\left\vert \left\{  \mathrm{H}\text{, }g\right\}  \right\vert \text{.}
\label{impo1}%
\end{equation}
However, Hamilton's equation of motion $dg/dt=\partial_{t}g+\left\{  g\text{,
}\mathrm{H}\right\}  $ reduces to $dg/dt=\left\{  g\text{, }\mathrm{H}%
\right\}  $ when $g$ does not exhibit any explicit time dependence. In this
working condition, Eq. (\ref{impo1}) reduces to%
\begin{equation}
\left(  \Delta E\right)  \left(  \Delta g\right)  \geq\frac{\hslash}%
{2}\left\vert \frac{dg}{dt}\right\vert \text{.} \label{landau}%
\end{equation}
Eq. (\ref{landau}) is the most useful inequality that appears in Landau's
theory of fluctuations where $\dot{g}\overset{\text{def}}{=}dg/dt$ is the
classical rate of change of the quantity $g$. As a conclusive remark, we note
that Eq. (\ref{landau}) requires the use of the Hamilton equation of motion
$dg/dt=\partial_{t}g+\left\{  g\text{, }\mathrm{H}\right\}  $. This, in turn,
can be viewed as a restatement of a generalized version of Ehrenfest's theorem
expressed as $d\left\langle \hat{g}\right\rangle /dt=\left\langle \partial
_{t}\hat{g}\right\rangle +\left\langle \left[  \hat{g}\text{, }\mathrm{\hat
{H}}\right]  \right\rangle /(i\hslash)$ \cite{sakurai85}. Eq. (\ref{landau})
plays a key role in Pati's rederivation of Caianiello's 1984 proposal, as we
shall see shortly.

\subsection{The derivation of the maximal acceleration}

Pati's derivation of the maximal acceleration in Ref. \cite{pati92b} is based
upon the relation in Eq. (\ref{landau}) along with a minimal position-momentum
uncertainty relation for any quantum particle. Assume that $g\left(  t\right)
$ in Eq. (\ref{landau}) is equal to $x\left(  t\right)  $ (or, $v\left(
t\right)  \overset{\text{def}}{=}\left\vert dx/dt\right\vert \geq0$), the
position coordinate of the particle (or, the instantaneous velocity of the
particle). Then, Eq. (\ref{landau}) yields%
\begin{equation}
\left(  \Delta E\right)  \left(  \Delta x\right)  \geq\frac{\hslash}%
{2}v\text{, and }\left(  \Delta E\right)  \left(  \Delta v\right)  \geq
\frac{\hslash}{2}a\text{,} \label{yo1a}%
\end{equation}
respectively, with $a\left(  t\right)  \overset{\text{def}}{=}\left\vert
dv/dt\right\vert \geq0$ being the instantaneous acceleration of the particle.
Since all quantities in Eq. (\ref{yo1a}) are positive, its manipulation leads
to%
\begin{equation}
a\leq\frac{4}{\hslash^{2}}\left(  \Delta E\right)  ^{2}\left(  \Delta
x\right)  \left(  \frac{\Delta v}{v}\right)  \text{.} \label{yo2a}%
\end{equation}
Noting that $\Delta E=\left\vert dE/dp\right\vert \cdot\Delta p=v\cdot\Delta
p$, Eq. (\ref{yo2a}) becomes%
\begin{equation}
a\leq\frac{4}{\hslash^{2}}\left(  \Delta p\right)  ^{2}\left(  \Delta
x\right)  \left(  \Delta v\right)  v\text{.} \label{yo3}%
\end{equation}
Observing that $v\left(  \Delta v\right)  \leq v_{\max}^{2}\leq c^{2}$ and
assuming $\left(  \Delta p\right)  \left(  \Delta x\right)  =\hslash/2$ (i.e.,
assuming the particle is in a state of minimum position-momentum uncertainty
relation), Eq. (\ref{yo3}) yields%
\begin{equation}
a\leq\frac{c^{2}}{\Delta x}\text{,}%
\end{equation}
that is,%
\begin{equation}
a_{\max}=\frac{c^{2}}{\Delta x}\text{.} \label{yo4}%
\end{equation}
Interestingly, observe that once one sets $a_{\max}=a_{\max}\left(  c\text{,
}\Delta x\right)  $, dimensional analysis leads naturally to Eq. (\ref{yo4}).
Since $a_{\max}$ depends on the uncertainty $\Delta x$, Eq. (\ref{yo4})
implies that the concept of maximal acceleration is merely a quantum concept.
Finally, by assuming that the uncertainty in position measurement $2\left(
\Delta x\right)  $ is always greater than the Compton wavelength of the
particle,%
\begin{equation}
2\left(  \Delta x\right)  \geq\lambda_{\mathrm{Compton}}\overset{\text{def}%
}{=}\hslash/\left(  m_{0}c\right)  \text{,}%
\end{equation}
with $m_{0}$ denoting the proper mass of the particle, Eq. (\ref{yo4}) finally
reduces to Caianiello's Eq. (\ref{cai3}), $a_{\mathrm{\max}}=2\left(
m_{0}c^{3}\right)  /\hslash$. In addition to addressing the criticisms moved
against Caianiello' 1984 proposal, Pati pointed out an intriguing link between
$a_{\max}$ and the maximal energy loss in Ref. \cite{pati92b}. This link will
be discussed in the next subsection and will serve to partially motivate our
line of reasoning in the last part of Section IV.

\subsection{Link between maximal energy loss and maximal acceleration}

Pati showed in Ref. \cite{pati92b} that the existence of a maximal
acceleration for a quantum particle implies the existence of an upper limit to
its power loss, i.e., to its energy radiated per unit time. In particular, he
showed that the power loss is proportional to the square of the maximal
acceleration. The proof goes as follows. Using Eq. (\ref{landau}), assume
$g\left(  t\right)  $ is the momentum $p\left(  t\right)  $ of the particle
with $F\overset{\text{def}}{=}\left\vert dp/dt\right\vert $ being the
magnitude of the force that is responsible for the acceleration. Then, using
Eq. (\ref{landau}), we get that $\left(  \Delta E\right)  \left(  \Delta
x\right)  \geq\left(  \hslash/2\right)  v$ with $v\overset{\text{def}%
}{=}\left\vert dx/dt\right\vert $ and $\left(  \Delta E\right)  \left(  \Delta
v\right)  \geq\left(  \hslash/2\right)  a$ with $a\overset{\text{def}%
}{=}\left\vert dv/dt\right\vert $. Then, imposing $\left(  \Delta p\right)
\left(  \Delta x\right)  =\hslash/2$ and defining with $\mathrm{P}%
\overset{\text{def}}{=}\left\vert dE/dt\right\vert =Fv$ the energy radiated
per unit time, one obtains \cite{pati92b}%
\begin{equation}
\mathrm{P}_{\max}=\frac{\hslash}{2}\frac{a_{\max}^{2}}{c^{2}}\text{,}
\label{power}%
\end{equation}
with $a_{\max}=c^{2}/\Delta x$. In summary, the maximal energy loss
$\mathrm{P}_{\max}$ of the quantum particle scales like the maximal
acceleration squared, $a_{\max}^{2}$.

We are now ready to reconsider Pati's suggested link between the maximal
acceleration of a quantum particle and the speed of transportation in
projective Hilbert space. This link is essential for our work in Section IV.

\subsection{Link between maximal acceleration and speed of transportation in
projective Hilbert space}

Let $a\overset{\text{def}}{=}\left\vert d_{t}\left\langle \psi\left\vert
v\right\vert \psi\right\rangle \right\vert \geq0$ with $d_{t}%
\overset{\text{def}}{=}d/dt$ denote the acceleration of a quantum particle in
a state $\left\vert \psi\right\rangle $. Note that $a=a\left(  t\right)  $ is
a positive scalar quantity. Let $v$ be the velocity operator of the particle
given by $v\overset{\text{def}}{=}dx/dt=\left[  x\text{, }\mathrm{H}\right]
/(i\hslash)$. Clearly, $x$ and $\mathrm{H}$ denote the position and the
Hamiltonian operators, respectively. In Ref. \cite{pati92}, Pati showed that
the maximal acceleration $a_{\max}$ of a quantum particle is limited by the
speed of transportation in projective Hilbert space $v_{\mathrm{H}%
}\overset{\text{def}}{=}\Delta\mathrm{H}/\hslash$ with $\Delta\mathrm{H}%
\overset{\text{def}}{=}\sqrt{\left\langle \left(  \mathrm{H}-\left\langle
\mathrm{H}\right\rangle \right)  ^{2}\right\rangle }$ being the energy
uncertainty of the system specified by the Hamiltonian $\mathrm{H}$. Observe
that the quantity $v_{\mathrm{H}}$ has physical dimensions of the reciprocal
of time\textbf{.} In what follows, we revisit Pati's proof. From quantum
mechanics, we have
\begin{equation}
\left(  \Delta v\right)  \left(  \Delta\mathrm{H}\right)  \geq\frac{1}%
{2}\left\vert \left\langle \psi\left\vert \left[  v\text{, }\mathrm{H}\right]
\right\vert \psi\right\rangle \right\vert \text{,} \label{pata1}%
\end{equation}
with $\left\vert \psi\right\rangle =\left\vert \psi\left(  t\right)
\right\rangle $. If the velocity operator does not depend on time in an
explicit manner, recalling the generalized version of the Ehrenfest theorem as
$\partial_{t}\left\langle v\right\rangle =\left\langle \partial_{t}%
v\right\rangle +\left\langle \left[  v\text{, }\mathrm{H}\right]
\right\rangle /(i\hslash)$, we have $a=\left\langle \left[  v\text{,
}\mathrm{H}\right]  \right\rangle /(i\hslash)$. Therefore, Eq. (\ref{pata1})
reduces to%
\begin{equation}
a\leq\frac{2}{\hslash}\left(  \Delta v\right)  \left(  \Delta\mathrm{H}%
\right)  \text{.} \label{pata2}%
\end{equation}
Since $\Delta v^{2}\overset{\text{def}}{=}\left\langle v^{2}\right\rangle
-\left\langle v\right\rangle ^{2}\leq v_{\max}^{2}\leq c^{2}$, we have $\Delta
v\leq v_{\mathrm{\max}}\leq c$. Therefore, from Eq. (\ref{pata2}) we arrive
at
\begin{equation}
a\leq2c\frac{\Delta\mathrm{H}}{\hslash}\text{.} \label{pata3}%
\end{equation}
However, from the study of the geometry of quantum evolutions \cite{anandan90}%
, $\Delta\mathrm{H}/\hslash$ equals the speed of transportation in projective
Hilbert space $v_{\mathrm{H}}$. We emphasize that $v_{\mathrm{H}}$ has a
geometric significance since its determination does not require a detailed
knowledge of the Hamiltonian of the system. We conclude that $a\leq
2cv_{\mathrm{H}}$ and, finally%
\begin{equation}
a_{\max}\left(  t\right)  =2cv_{\mathrm{H}}\left(  t\right)  \text{.}
\label{pata4}%
\end{equation}
The derivation of Eq. (\ref{pata4}) ends our revisitation. Given the geometric
meaning of $v_{\mathrm{H}}$, Eq. (\ref{pata4}) implies that $a_{\max}$ does
not depend on the details of the generator of time translations $\mathrm{H}$
and is a geometric property of the evolution of the quantum system.

\section{Upper bound on the acceleration in projective Hilbert space}

We begin here by providing our novel upper bound on the acceleration in
projective Hilbert space in terms of the square-root of the dispersion of the
time-derivative of the nonstationary Hamiltonian of the finite-dimensional
quantum system under consideration. We then conclude with a practical
discussion on the consequences of our upper bound for the maximal jerk of a
spin-$1/2$ particle viewed as a single-qubit quantum system. However, before
discussing the upper bound, we present some background information that
originally motivated us to consider the inequality leading to the newly
proposed upper bound.

\subsection{Background motivation}

In Ref. \cite{laba17}, Laba and Tkachuk proposed a unitless and positively
constant curvature coefficient $\kappa_{\mathrm{LT}}^{2}$ for quantum
evolutions of finite-dimensional systems specified by stationary Hamiltonians
\textrm{H},%
\begin{equation}
\kappa_{\mathrm{LT}}^{2}=\left\langle \left(  \Delta h\right)  ^{4}%
\right\rangle -\left\langle \left(  \Delta h\right)  ^{2}\right\rangle
^{2}\text{.} \label{LT}%
\end{equation}
Eq. (\ref{LT}) was recently extended to unitless and positively nonconstant
curvature coefficients $\kappa_{\mathrm{AC}}^{2}$ for quantum evolutions of
finite-dimensional systems characterized by nonstationary Hamiltonians
\textrm{H}$\left(  t\right)  $ by Alsing and Cafaro in Refs.
\cite{alsing1,alsing2},%
\begin{equation}
\kappa_{\mathrm{AC}}^{2}=\left\langle \left(  \Delta h\right)  ^{4}%
\right\rangle -\left\langle \left(  \Delta h\right)  ^{2}\right\rangle
^{2}+\left[  \left\langle \left(  \Delta h^{\prime}\right)  ^{2}\right\rangle
-\left\langle \Delta h^{\prime}\right\rangle ^{2}\right]  +i\left\langle
\left[  \left(  \Delta h\right)  ^{2}\text{, }\Delta h^{\prime}\right]
\right\rangle \text{.} \label{AC}%
\end{equation}
In Eq. (\ref{AC}), $\Delta h$ is the unitless operator defined as $\Delta
h\overset{\text{def}}{=}\Delta\mathrm{H}/\left(  \hslash v_{\mathrm{H}%
}\right)  $ with $\Delta\mathrm{H}\overset{\text{def}}{=}\mathrm{H}%
-\left\langle \mathrm{H}\right\rangle $, and $v_{\mathrm{H}}$ being the speed
of evolution in projective Hilbert space. The speed $v_{\mathrm{H}}$ is given
by $v_{\mathrm{H}}\overset{\text{def}}{=}\sigma_{\mathrm{H}}/\hslash$, with
$\sigma_{\mathrm{H}}^{2}\overset{\text{def}}{=}\left\langle \left(
\Delta\mathrm{H}\right)  ^{2}\right\rangle $ being the variance of the
Hamiltonian operator \textrm{H}. Expectation values are to be taken for a
certain physical state being considered. The prime symbol in $\Delta
h^{\prime}$ denotes the differentiation with respect to the arc length
parameter $s$, $\Delta h^{\prime}\overset{\text{def}}{=}\partial_{s}\left(
\Delta h\right)  $ with $s\left(  t\right)  \overset{\text{def}}{=}\int%
_{0}^{t}v_{\mathrm{H}}\left(  \tau\right)  d\tau$. The square brackets in
$\left[  \left(  \Delta h\right)  ^{2}\text{, }\Delta h^{\prime}\right]  $
denote the quantum commutator and, $i$ is the usual complex imaginary unit.
While focusing on the interpretation and signs of the various terms in
$\kappa_{\mathrm{AC}}^{2}$, one notices that the term $\left\langle \left(
\Delta h^{\prime}\right)  ^{2}\right\rangle -\left\langle \Delta h^{\prime
}\right\rangle ^{2}$ reduces to $\left\langle \left(  \Delta h^{\prime
}\right)  ^{2}\right\rangle $ since $\left\langle \Delta h^{\prime
}\right\rangle $ is identically zero. This can be explicitly checked by
jointly using the definition $\Delta h\overset{\text{def}}{=}\Delta
\mathrm{H}/\left(  \hslash v_{\mathrm{H}}\right)  $ and the differentiation
rule $\partial_{s}\overset{\text{def}}{=}v_{\mathrm{H}}^{-1}\partial_{t}$.
Therefore, one recognizes the positivity of $\left\langle \left(  \Delta
h^{\prime}\right)  ^{2}\right\rangle $ which, in turn, can be conveniently
recast as%
\begin{equation}
\left\langle \left(  \Delta h^{\prime}\right)  ^{2}\right\rangle
=\frac{\left\langle \left(  \Delta\mathrm{\dot{H}}\right)  ^{2}\right\rangle
-\left(  \partial_{t}\sqrt{\left\langle \left(  \Delta\mathrm{H}\right)
^{2}\right\rangle }\right)  ^{2}}{\left\langle \left(  \Delta\mathrm{H}%
\right)  ^{2}\right\rangle ^{2}}\text{,} \label{ac1}%
\end{equation}
with $\Delta\mathrm{\dot{H}}\overset{\text{def}}{=}\partial_{t}\left(
\Delta\mathrm{H}\right)  =\mathrm{\dot{H}}-\left\langle \mathrm{\dot{H}%
}\right\rangle $ and $\mathrm{\dot{H}}\overset{\text{def}}{=}\partial
_{t}\mathrm{H}$. Finally, we arrive from Eq. (\ref{ac1}) at the inequality
that we wish to prove%
\begin{equation}
a_{\mathrm{H}}^{2}\leq\sigma_{\mathrm{\dot{H}}}^{2}\text{.} \label{TI}%
\end{equation}
In Eq. (\ref{TI}), $a_{\mathrm{H}}\overset{\text{def}}{=}\partial
_{t}v_{\mathrm{H}}=\partial_{t}\sigma_{\mathrm{H}}=\dot{\sigma}_{\mathrm{H}}$
is the acceleration of the quantum evolution in projective Hilbert space,
while $\sigma_{\mathrm{\dot{H}}}^{2}=\sigma_{\partial_{t}\mathrm{H}}%
^{2}=\left\langle \mathrm{\dot{H}}^{2}\right\rangle -\left\langle
\mathrm{\dot{H}}\right\rangle ^{2}$ is the dispersion of the time-derivative
$\mathrm{\dot{H}}$ of the nonstationary Hamiltonian operator $\mathrm{H}%
=\mathrm{H}(t)$.

\subsection{Upper bound: General case}

We wish to prove that for any finite-dimensional quantum system that evolves
under the time-dependent Hamiltonian \textrm{H}$(t)$, we have%
\begin{equation}
\left(  \partial_{t}\sigma_{\mathrm{H}}\right)  ^{2}\leq\left(  \sigma
_{\partial_{t}\mathrm{H}}\right)  ^{2}\text{,} \label{one}%
\end{equation}
that is, Eq. (\ref{TI}) with $a_{\mathrm{H}}\overset{\text{def}}{=}%
\partial_{t}\sigma_{\mathrm{H}}$ and $\sigma_{\mathrm{\dot{H}}}%
\overset{\text{def}}{=}\sigma_{\partial_{t}\mathrm{H}}$. We shall show that
the inequality in Eq. (\ref{one}) is a consequence of the generalized
uncertainty principle in quantum theory. Following Ref. \cite{sakurai85}, the
generalized uncertainty principle states that any pair of quantum observables
$A$ and $B$ satisfy the inequality%
\begin{equation}
\left\langle \left(  \Delta A\right)  ^{2}\right\rangle \left\langle \left(
\Delta B\right)  ^{2}\right\rangle \geq\frac{1}{4}\left\vert \left\langle
\left[  A\text{, }B\right]  \right\rangle \right\vert ^{2}\text{,}
\label{sakurai}%
\end{equation}
where the expectation values in Eq. (\ref{sakurai}) are calculated with
respect to a given physical state. We begin by observing that since $\left(
\partial_{t}\sigma_{\mathrm{H}}\right)  ^{2}=\left(  \partial_{t}\left\langle
\left(  \Delta\mathrm{H}\right)  ^{2}\right\rangle \right)  ^{2}/4\left\langle
\left(  \Delta\mathrm{H}\right)  ^{2}\right\rangle $ and $\left(
\sigma_{\partial_{t}\mathrm{H}}\right)  ^{2}=\left\langle \left(
\Delta\mathrm{\dot{H}}\right)  ^{2}\right\rangle $, Eq. (\ref{one}) can be
recast as%
\begin{equation}
\left\langle \left(  \Delta\mathrm{\dot{H}}\right)  ^{2}\right\rangle
\left\langle \left(  \Delta\mathrm{H}\right)  ^{2}\right\rangle \geq\frac
{1}{4}\left[  \partial_{t}\left\langle \left(  \Delta\mathrm{H}\right)
^{2}\right\rangle \right]  ^{2}\text{.} \label{love1}%
\end{equation}
At this point, we note that the term $\left[  \partial_{t}\left\langle \left(
\Delta\mathrm{H}\right)  ^{2}\right\rangle \right]  ^{2}$ in Eq. (\ref{love1})
can be conveniently expressed in terms of a quantum anti-commutator,%
\begin{equation}
\left[  \partial_{t}\left\langle \left(  \Delta\mathrm{H}\right)
^{2}\right\rangle \right]  ^{2}=\left[  \left\langle \left(  \Delta
\mathrm{\dot{H}}\right)  \left(  \Delta\mathrm{H}\right)  +\left(
\Delta\mathrm{H}\right)  \left(  \Delta\mathrm{\dot{H}}\right)  \right\rangle
\right]  ^{2}=\left\langle \left\{  \Delta\mathrm{\dot{H}}\text{, }%
\Delta\mathrm{H}\right\}  \right\rangle ^{2}\text{.} \label{love2}%
\end{equation}
Therefore, combing Eqs. (\ref{love1}) and (\ref{love2}), the inequality
in\ Eq. (\ref{one}) becomes%
\begin{equation}
\left\langle \left(  \Delta\mathrm{\dot{H}}\right)  ^{2}\right\rangle
\left\langle \left(  \Delta\mathrm{H}\right)  ^{2}\right\rangle \geq\frac
{1}{4}\left\langle \left\{  \Delta\mathrm{\dot{H}}\text{, }\Delta
\mathrm{H}\right\}  \right\rangle ^{2}\text{.} \label{b}%
\end{equation}
We note from Eq. (\ref{b}) that $\left\langle \left(  \Delta\mathrm{\dot{H}%
}\right)  ^{2}\right\rangle \left\langle \left(  \Delta\mathrm{H}\right)
^{2}\right\rangle =\left\langle \Delta\mathrm{\dot{H}}\left\vert
\Delta\mathrm{\dot{H}}\right.  \right\rangle \left\langle \Delta
\mathrm{H}\left\vert \Delta\mathrm{H}\right.  \right\rangle $. Then, using the
Schwarz inequality \cite{sakurai85}, we have%
\begin{equation}
\left\langle \Delta\mathrm{\dot{H}}\left\vert \Delta\mathrm{\dot{H}}\right.
\right\rangle \left\langle \Delta\mathrm{H}\left\vert \Delta\mathrm{H}\right.
\right\rangle \geq\left\vert \left\langle \left(  \Delta\mathrm{H}\right)
\left(  \Delta\mathrm{\dot{H}}\right)  \right\rangle \right\vert ^{2}\text{,}%
\end{equation}
that is,%
\begin{equation}
\left\langle \left(  \Delta\mathrm{\dot{H}}\right)  ^{2}\right\rangle
\left\langle \left(  \Delta\mathrm{H}\right)  ^{2}\right\rangle \geq\left\vert
\left\langle \left(  \Delta\mathrm{H}\right)  \left(  \Delta\mathrm{\dot{H}%
}\right)  \right\rangle \right\vert ^{2}\text{.} \label{b4}%
\end{equation}
From Eqs. (\ref{b}) and (\ref{b4}), we conclude that the inequality in Eq.
(\ref{b}) can be proved if we are able to verify that
\begin{equation}
\left\vert \left\langle \left(  \Delta\mathrm{H}\right)  \left(
\Delta\mathrm{\dot{H}}\right)  \right\rangle \right\vert ^{2}\geq\frac{1}%
{4}\left\langle \left\{  \Delta\mathrm{\dot{H}}\text{, }\Delta\mathrm{H}%
\right\}  \right\rangle ^{2}\text{.} \label{c}%
\end{equation}
Therefore, let us focus on proving the inequality in Eq. (\ref{c}). Note that
$2\left(  \Delta\mathrm{H}\right)  \left(  \Delta\mathrm{\dot{H}}\right)  $
can be decomposed as a sum of a commutator and an anti-commutator,%
\begin{equation}
2\left(  \Delta\mathrm{H}\right)  \left(  \Delta\mathrm{\dot{H}}\right)
=\left[  \Delta\mathrm{H}\text{, }\Delta\mathrm{\dot{H}}\right]  +\left\{
\Delta\mathrm{H}\text{, }\Delta\mathrm{\dot{H}}\right\}  =\left[
\Delta\mathrm{H}\text{, }\Delta\mathrm{\dot{H}}\right]  +\left\{
\Delta\mathrm{\dot{H}}\text{, }\Delta\mathrm{H}\right\}  \text{.} \label{d1}%
\end{equation}
Given that $\left[  \Delta\mathrm{H}\text{, }\Delta\mathrm{\dot{H}}\right]  $
is an anti-Hermitian operator, $\ \left\langle \left[  \Delta\mathrm{H}\text{,
}\Delta\mathrm{\dot{H}}\right]  \right\rangle $ is purely imaginary.
Furthermore, given that $\left\{  \Delta\mathrm{\dot{H}}\text{, }%
\Delta\mathrm{H}\right\}  $ is a Hermitian operator, its expectation value
$\left\langle \left\{  \Delta\mathrm{\dot{H}}\text{, }\Delta\mathrm{H}%
\right\}  \right\rangle $ is real. Therefore, exploiting these considerations,
we obtain from Eq. (\ref{d1}) that%
\begin{equation}
4\left\vert \left\langle \left(  \Delta\mathrm{H}\right)  \left(
\Delta\mathrm{\dot{H}}\right)  \right\rangle \right\vert ^{2}=\left\vert
\left\langle \left[  \Delta\mathrm{H}\text{, }\Delta\mathrm{\dot{H}}\right]
\right\rangle \right\vert ^{2}+\left\vert \left\langle \left\{  \Delta
\mathrm{\dot{H}}\text{, }\Delta\mathrm{H}\right\}  \right\rangle \right\vert
^{2}\text{,}%
\end{equation}
that is,%
\begin{equation}
\left\vert \left\langle \left(  \Delta\mathrm{H}\right)  \left(
\Delta\mathrm{\dot{H}}\right)  \right\rangle \right\vert ^{2}=\frac{\left\vert
\left\langle \left[  \Delta\mathrm{H}\text{, }\Delta\mathrm{\dot{H}}\right]
\right\rangle \right\vert ^{2}+\left\vert \left\langle \left\{  \Delta
\mathrm{\dot{H}}\text{, }\Delta\mathrm{H}\right\}  \right\rangle \right\vert
^{2}}{4}\geq\frac{1}{4}\left\vert \left\langle \left\{  \Delta\mathrm{\dot{H}%
}\text{, }\Delta\mathrm{H}\right\}  \right\rangle \right\vert ^{2}\text{.}
\label{d}%
\end{equation}
We conclude that the inequality in Eq. (\ref{c}) is satisfied and, thus, our
inequality in Eq. (\ref{b}) is proved as well. In summary, the inequality
$\left(  \partial_{t}\sigma_{\mathrm{H}}\right)  ^{2}\leq$ $\left(
\sigma_{\partial_{t}\mathrm{H}}\right)  ^{2}\Leftrightarrow a_{\mathrm{H}}%
^{2}\leq\sigma_{\mathrm{\dot{H}}}^{2}$ has foundational relevance and is of
general validity for any nonstationary Hamiltonian evolution. The inequality
$a_{\mathrm{H}}^{2}\leq\sigma_{\mathrm{\dot{H}}}^{2}$ implies that
$a_{\mathrm{H}}\leq\left\vert a_{\mathrm{H}}\right\vert \leq\sigma
_{\mathrm{\dot{H}}}$, that is, the acceleration of transportation in
projective Hilbert space is upper bounded by the square root of the dispersion
of the time-derivative of the Hamiltonian operator. Interestingly, we get from
Eq. (\ref{pata4}) that%
\begin{equation}
\dot{a}_{\max}\left(  t\right)  =2c\cdot\dot{v}_{\mathrm{H}}\left(  t\right)
=2c\cdot a_{\mathrm{H}}\left(  t\right)  \leq2c\cdot\left\vert a_{\mathrm{H}%
}\left(  t\right)  \right\vert \leq2c\cdot\sigma_{\mathrm{\dot{H}}}\left(
t\right)  \text{,}%
\end{equation}
that is,
\begin{equation}
\dot{a}_{\max}\left(  t\right)  \leq2c\cdot\sigma_{\mathrm{\dot{H}}}\left(
t\right)  \text{.} \label{jerky}%
\end{equation}
Eq. (\ref{jerky}) expresses the fact that the maximal jerk $\bar{j}\left(
t\right)  \overset{\text{def}}{=}$ $\dot{a}_{\max}\left(  t\right)  $ of a
quantum system is, modulo a constant factor that equals $2c$, essentially
upper bounded by the square root of the dispersion of the time-derivative of
the Hamiltonian of the system under investigation. For illustrative purposes,
we shall discuss the consequences of Eqs. (\ref{one}) and (\ref{jerky}) on the
maximal jerk of a spin-$1/2$ quantum particle in the next subsection.

\subsection{Upper bound: Single-qubit case}

Consider a physical system specified by the density operator $\rho\left(
t\right)  \overset{\text{def}}{=}\left[  \mathrm{I}+\mathbf{a}\left(
t\right)  \mathbf{\cdot}\vec{\sigma}\right]  /2$ that evolves under the
time-dependent Hamiltonian \textrm{H}$\left(  t\right)  \overset{\text{def}%
}{=}\mathbf{m}\left(  t\right)  \cdot\vec{\sigma}$. The vectors $\mathbf{a}$
and $\mathbf{m}\overset{\text{def}}{=}\left\vert e\right\vert \mathbf{B}%
/(2m_{e}c)$ denote the Bloch vector and the magnetic vector, respectively.
Clearly, $m_{e}$ is the mass of an electron and $\mathbf{B}$ denotes the
magnetic field vector. Interestingly, an arbitrary qubit observable
$Q=q_{0}\mathrm{I}+\mathbf{q\cdot}\vec{\sigma}$ with $q_{0}\in%
\mathbb{R}
$ and $\mathbf{q\in%
\mathbb{R}
}^{3}$ has its expectation value $\left\langle Q\right\rangle _{\rho}%
=q_{0}+\mathbf{a\cdot q}$. For simplicity, we set $\hslash=1$ in what follows.
Then, the speed of the quantum evolution in projective Hilbert space equals
$v_{\mathrm{H}}\overset{\text{def}}{=}\Delta\mathrm{H}$. The quantity
$\Delta\mathrm{H}$ is the standard deviation $\sigma_{\mathrm{H}}$ of the
Hamiltonian \textrm{H}, with $\sigma_{\mathrm{H}}\overset{\text{def}%
}{=}\left[  \left\langle \mathrm{H}^{2}\right\rangle -\left\langle
\mathrm{H}\right\rangle ^{2}\right]  ^{1/2}$. The acceleration of the quantum
evolution in projective Hilbert space, instead, equals $a_{\mathrm{H}%
}\overset{\text{def}}{=}\dot{v}_{\mathrm{H}}$ with $\dot{v}_{\mathrm{H}%
}=\partial_{t}v_{\mathrm{H}}$. In terms of the Bloch and magnetic vectors,
$v_{\mathrm{H}}$ and $a_{\mathrm{H}}$ assume the expressions
\ \ \ \ \ \ \ \ \ \
\begin{equation}
v_{\mathrm{H}}=\sqrt{\mathbf{m}^{2}-\left(  \mathbf{a\cdot m}\right)  ^{2}%
}\text{, and }a_{\mathrm{H}}=\frac{\mathbf{m\cdot\dot{m}-}\left(
\mathbf{a\cdot m}\right)  \left[  \partial_{t}\left(  \mathbf{a\cdot
m}\right)  \right]  }{\sqrt{\mathbf{m}^{2}-\left(  \mathbf{a\cdot m}\right)
^{2}}}\text{,} \label{got1}%
\end{equation}
respectively, since $\left\langle \mathrm{H}\right\rangle =\mathbf{a\cdot m}$
and $\left\langle \mathrm{H}^{2}\right\rangle =\mathbf{m}^{2}$. From the first
relation in Eq. (\ref{got1}), we observe that the orthogonality between the
Bloch vector $\mathbf{a}$ and the magnetic vector $\mathbf{m}$ is necessary to
achieve maximal dispersion of the Hamiltonian operator. When $\mathbf{a\perp
m}$, $v_{\mathrm{H}}$ assumes its maximal value of $\left(  v_{\mathrm{H}%
}\right)  _{\max}=\left\Vert \mathbf{m}\right\Vert $. From the second relation
in Eq. (\ref{got1}), we note that the acceleration $a_{\mathrm{H}}$ assumes
its maximal value of%
\begin{equation}
\left(  a_{\mathrm{H}}\right)  _{\max}=\left\Vert \mathbf{\dot{m}}\right\Vert
\text{,} \label{maxa}%
\end{equation}
when $\mathbf{a\cdot m}=0$ and, in addition, $\mathbf{m}$ and $\mathbf{\dot
{m}}$ are collinear. However, the collinearity between $\mathbf{m}$ and
$\mathbf{\dot{m}}$ is not really an additional condition to be satisfied.
Indeed, since $\mathbf{\dot{a}=}2\mathbf{m\times a}$ (for details, see
Appendix A), $\mathbf{\dot{a}\perp m}$. Therefore, when $\mathbf{a\perp m}$,
one has $0=\partial_{t}\left(  \mathbf{a\cdot m}\right)  =\mathbf{\dot{a}\cdot
m+a\cdot}$ $\mathbf{\dot{m}=a\cdot}$ $\mathbf{\dot{m}}$. Thus, $\mathbf{a\perp
\dot{m}}$ when $\mathbf{a\perp m}$. Since $\mathbf{\dot{a}}$, $\mathbf{a}$,
$\mathbf{m}$, and $\mathbf{\dot{m}}$ are all three-dimensional vectors, we
conclude from our previous observations the following: Anytime $\mathbf{a\perp
m}$, the sets $\left\{  \mathbf{\dot{a}},\mathbf{a},\mathbf{m}\right\}  $ and
$\left\{  \mathbf{\dot{a}},\mathbf{a},\mathbf{\dot{m}}\right\}  $ are sets of
orthogonal three-dimensional vectors with $\mathbf{m}$ and $\mathbf{\dot{m}}$
necessarily collinear. In summary, the collinearity of $\mathbf{m}$ and
$\mathbf{\dot{m}}$ can be viewed as a consequence of the relation
$\mathbf{\dot{a}=}2\mathbf{m\times a}$ along with the assumptions of
orthogonality between $\mathbf{a}$ and $\mathbf{m}$. As a side remark, note
that in general $\mathbf{m=m}\left(  t\right)  \overset{\text{def}}{=}%
m(t)\hat{m}(t)$. Therefore, $\mathbf{\dot{m}=\dot{m}}\left(  t\right)
\overset{\text{def}}{=}\dot{m}(t)\hat{m}(t)+m(t)\left[  \partial_{t}\hat
{m}(t)\right]  $ which, in turn, implies that $\mathbf{m}$ and $\mathbf{\dot
{m}}$ are collinear when $\partial_{t}\hat{m}(t)$ is zero and the magnetic
vector $\mathbf{m}$ changes only in its intensity.

The necessity of the collinearity between $\mathbf{m}$ and $\mathbf{\dot{m}}$
in order to achieve maximal values of the acceleration $a_{\mathrm{H}}$ is
clear when considering the qubit-version of the inequality in Eq. (\ref{one}),
$\left(  \partial_{t}\sigma_{\mathrm{H}}\right)  ^{2}\leq\sigma_{\partial
_{t}\mathrm{H}}^{2}$. Since $\partial_{t}\sigma_{\mathrm{H}}=\dot
{v}_{\mathrm{H}}=a_{\mathrm{H}}$ and $\sigma_{\partial_{t}\mathrm{H}}%
^{2}=\left\langle (\mathrm{\dot{H}})^{2}\right\rangle -\left\langle
\mathrm{\dot{H}}\right\rangle ^{2}$, the inequality $\left(  \partial
_{t}\sigma_{\mathrm{H}}\right)  ^{2}\leq\sigma_{\partial_{t}\mathrm{H}}^{2}$
becomes $a_{\mathrm{H}}^{2}\leq\left\langle (\mathrm{\dot{H}})^{2}%
\right\rangle -\left\langle \mathrm{\dot{H}}\right\rangle ^{2}$. Then, noting
that $\left\langle (\mathrm{\dot{H}})^{2}\right\rangle =\mathbf{\dot{m}}^{2}$
and $\left\langle \mathrm{\dot{H}}\right\rangle =\partial_{t}\left(
\mathbf{a\cdot m}\right)  \mathbf{=a\cdot}$ $\mathbf{\dot{m}}$ since
$\mathbf{\dot{a}=}2\mathbf{m\times a}$, $\left(  \partial_{t}\sigma
_{\mathrm{H}}\right)  ^{2}\leq\sigma_{\partial_{t}\mathrm{H}}^{2}$ reduces to%
\begin{equation}
a_{\mathrm{H}}^{2}\leq\mathbf{\dot{m}}^{2}-\left(  \mathbf{a\cdot\dot{m}%
}\right)  ^{2}\text{.} \label{ineq2}%
\end{equation}
Once again, given that $\mathbf{\dot{a}=}2\mathbf{m\times a}$ and assuming the
orthogonality between $\mathbf{a}$ and $\mathbf{m}$, the maximal value of
$a_{\mathrm{H}}$ becomes $\left(  a_{\mathrm{H}}\right)  _{\max}=\left\Vert
\mathbf{\dot{m}}\right\Vert $. This is in agreement with the result reported
in Eq. (\ref{maxa}). Interestingly, the inequality in Eq. (\ref{one}) can be
fully expressed in terms of $\mathbf{a}$, $\mathbf{m}$, and $\mathbf{\dot{m}}%
$. After some algebraic manipulations, we get
\begin{equation}
\left[  \mathbf{m}^{2}\mathbf{\dot{m}}^{2}-\left(  \mathbf{m\cdot\dot{m}%
}\right)  ^{2}\right]  \geq\left[  \left(  \mathbf{a\cdot\dot{m}}\right)
\mathbf{m-}\left(  \mathbf{a\cdot m}\right)  \mathbf{\dot{m}}\right]
^{2}\text{.} \label{cul4}%
\end{equation}
Finally, exploiting the Lagrange identity and the vector triple product
relations (i. e., $\left\Vert \mathbf{v}_{1}\times\mathbf{v}_{2}\right\Vert
^{2}=\left(  \mathbf{v}_{1}\cdot\mathbf{v}_{1}\right)  \left(  \mathbf{v}%
_{2}\cdot\mathbf{v}_{2}\right)  -\left(  \mathbf{v}_{1}\cdot\mathbf{v}%
_{2}\right)  ^{2}$ and $\mathbf{v}_{1}\times\left(  \mathbf{v}_{2}%
\times\mathbf{v}_{3}\right)  =\left(  \mathbf{v}_{1}\cdot\mathbf{v}%
_{3}\right)  \mathbf{v}_{2}-\left(  \mathbf{v}_{1}\cdot\mathbf{v}_{2}\right)
\mathbf{v}_{3}$, respectively), we arrive at the conclusion that
$\mathbf{m}^{2}\mathbf{\dot{m}}^{2}-\left(  \mathbf{m\cdot\dot{m}}\right)
^{2}=\left\Vert \mathbf{m}\times\mathbf{\dot{m}}\right\Vert ^{2}$ and $\left[
\left(  \mathbf{a\cdot\dot{m}}\right)  \mathbf{m-}\left(  \mathbf{a\cdot
m}\right)  \mathbf{\dot{m}}\right]  ^{2}=\left\Vert \mathbf{a\times}\left(
\mathbf{m}\times\mathbf{\dot{m}}\right)  \right\Vert ^{2}=\left(
\mathbf{a\cdot a}\right)  \left\Vert \mathbf{m}\times\mathbf{\dot{m}%
}\right\Vert ^{2}-\left[  \mathbf{a\cdot}\left(  \mathbf{m\times\dot{m}%
}\right)  \right]  ^{2}=\left\Vert \mathbf{m}\times\mathbf{\dot{m}}\right\Vert
^{2}-\left[  \mathbf{a\cdot}\left(  \mathbf{m\times\dot{m}}\right)  \right]
^{2}$ since $\mathbf{a\cdot a=}1$. Therefore, the inequality in Eq.
(\ref{cul4}) reduces to%
\begin{equation}
\left\Vert \mathbf{m}\times\mathbf{\dot{m}}\right\Vert ^{2}\geq\left\Vert
\mathbf{m}\times\mathbf{\dot{m}}\right\Vert ^{2}-\left[  \mathbf{a\cdot
}\left(  \mathbf{m\times\dot{m}}\right)  \right]  ^{2}\text{.} \label{cul5}%
\end{equation}
The inequality in Eq. (\ref{cul5}) is clearly satisfied for arbitrary vectors
$\mathbf{a}$, $\mathbf{m}$, and $\mathbf{\dot{m}}$. In particular, the
equality is achieved only when $\mathbf{a\cdot}\left(  \mathbf{m\times\dot{m}%
}\right)  =0$. This, in turn, happens if and only if $\sin\left(
\theta_{\mathbf{m,\dot{m}}}\right)  \cos\left(  \theta_{\mathbf{a,m}%
\times\mathbf{\dot{m}}}\right)  =0$, that is, if $\theta_{\mathbf{m,\dot{m}}%
}=k\pi$ or $\theta_{\mathbf{a,m}\times\mathbf{\dot{m}}}=(\pi/2)+k\pi$ with
$k\in%
\mathbb{Z}
$. Therefore, the necessity of the collinearity between $\mathbf{m}$ and
$\mathbf{\dot{m}}$ in order to achieve maximal values of the acceleration
$a_{\mathrm{H}}$ can be comprehended in these more geometric terms as well.
Two last remarks are in order. First, the average energy loss per unit time
$\left\langle \mathrm{\dot{H}}\right\rangle =\left\langle \partial
_{t}\mathrm{H}\right\rangle =\partial_{t}\left(  \mathbf{a\cdot m}\right)  $
is null when $\mathbf{a\cdot m}$ does not depend on time, that is, when
$\mathbf{a\perp\dot{m}}$ since $\mathbf{\dot{a}=}2\mathbf{m\times a}$.
Therefore, we can conclude that $\left\langle \partial_{t}\mathrm{H}%
\right\rangle $ vanishes when the speed and the acceleration in projective
Hilbert space assume their maximal values $\left\Vert \mathbf{m}\right\Vert $
and $\left\Vert \mathbf{\dot{m}}\right\Vert $, respectively. Second,
time-differentiating Pati's relation $a_{\max}\left(  t\right)
=2cv_{\mathrm{H}}\left(  t\right)  $, we get $\dot{a}_{\max}\left(  t\right)
=2c\dot{v}_{\mathrm{H}}\left(  t\right)  =2ca_{\mathrm{H}}\left(  t\right)  $.
Then, using Eq. (\ref{ineq2}) we get an upper bound on the maximal jerk of the
quantum particle (i.e., the electron in our case) given by%
\begin{equation}
\dot{a}_{\max}\left(  t\right)  \leq2c\left\Vert \mathbf{\dot{m}}\left(
t\right)  \right\Vert \text{.} \label{jerco}%
\end{equation}
Eq. (\ref{jerco}) implies that to minimize the maximal jerk of the particle
and to have a smooth quantum driving path, one needs to avoid highly intense
sudden changes in the magnetic field vector $\mathbf{m}\left(  t\right)  $
used in the quantum driving scheme specified by the Hamiltonian \textrm{H}%
$\left(  t\right)  \overset{\text{def}}{=}\mathbf{m}\left(  t\right)
\cdot\vec{\sigma}$. Moreover, we note that when $\mathrm{\dot{H}}$ vanishes,
$v_{\mathrm{H}}$ is constant and $\dot{a}_{\max}$ vanishes. In conclusion,
maximization of the speed of evolution in projective Hilbert space along with
minimal average energy loss per unit time, demands a magnetic field vector
$\mathbf{m}\left(  t\right)  =m(t)\hat{m}(t)$ that changes only in intensity
and not in direction. Furthermore, the smoothness of the path of evolution
requires the minimization of the jerk of the spin-$1/2$ particle being driven.
This is achieved by avoiding high-magnitude sudden changes in the external
magnetic field intensity and cleverly controlling $\dot{m}(t)$.

\section{Concluding remarks \ \ }

In this paper, we first derived an upper bound for the rate of change of the
speed (i.e., the acceleration) of transportation in an arbitrary
finite-dimensional projective Hilbert space (Eqs. (\ref{TI}) and (\ref{one})).
The evolution of the physical system being in a pure quantum state is assumed
to be governed by an arbitrary time-varying Hermitian Hamiltonian operator
\textrm{H}$=$\textrm{H}$\left(  t\right)  $. Our derivation, in analogy to the
inequalities obtained by L. D. Landau in the theory of fluctuations
(Eq.\ (\ref{landau})) by means of general commutation relations of
quantum-mechanical origin (Eqs. (\ref{impo}) and (\ref{impo1})), relies upon a
generalization of Heisenberg's uncertainty relation (Eq. (\ref{sakurai})).
More specifically, we showed in our derivation that the (modulus) of the
acceleration of a quantum evolution in projective space is upper bounded by
the square root of the dispersion of the temporal rate of change of the
Hamiltonian operator, $\left\vert a_{\mathrm{H}}\left(  t\right)  \right\vert
\leq\sigma_{\mathrm{\dot{H}}}$. We then used this inequality together with
Pati's Eq. (\ref{pata4}), $a_{\max}\left(  t\right)  =2c\cdot v_{\mathrm{H}%
}\left(  t\right)  $, to provide an upper bound on the maximal jerk of a
quantum system in terms of the standard deviation of the temporal rate of
change of the Hamiltonian operator (Eq. (\ref{jerky})). For illustrative
purposes,we then focused on the lower-dimensional case of a single spin-$1/2$
qubit immersed in an arbitrarily time-varying magnetic field and we discussed
the optimal geometric configuration of the magnetic field that yields maximal
acceleration along with vanishing curvature and unit geodesic efficiency in
projective Hilbert space. Finally, we commented on the consequences that our
upper bound imposes on the limit at which one can perform fast manipulations
of quantum systems to mitigate dissipative effects and/or obtain a target
state in a shorter time.

Despite the fact that our work is originally motivated by fundamental
interests, we recognize that its significance can have more applied
perspectives as well. Indeed, bounds on the speed, the acceleration, and the
jerk of a quantum evolution can be exploited to characterize the complexity
and efficiency of quantum processes
\cite{uzdin12,campaioli19,cafaroQR,carloprd22,carlopre22,carlocqg23}, to
estimate the cost of controlled quantum evolutions
\cite{frey16,deffner17,saito23}, and to quantify losses in the presence of
quantum critical behaviors \cite{liu19,matt23,araki23}. We leave the
investigation of some of these more applied and practical aspects to future
scientific efforts.\bigskip

\emph{Note}. After completion of this work as originally appeared in Ref.
\cite{carlo23}, we learned about Ref. \cite{pati23} devoted to the existence
of a quantum acceleration limit in projective Hilbert space.

\begin{acknowledgments}
P.M.A. acknowledges support from the Air Force Office of Scientific Research
(AFOSR). C.C. is grateful to the United States Air Force Research Laboratory
(AFRL) Summer Faculty Fellowship Program for providing support for this work.
Any opinions, findings and conclusions or recommendations expressed in this
material are those of the author(s) and do not necessarily reflect the views
of the Air Force Research Laboratory (AFRL).
\end{acknowledgments}

\pagebreak

\appendix

\section{Derivation of $\mathbf{\dot{a}=}2\mathbf{m\times a}$}

In this Appendix, we show the relation $\mathbf{\dot{a}=}2\mathbf{m\times a}$
that appears in subsection C of Section IV. Assume that \textrm{H}$\left(
t\right)  =\mathbf{m}\left(  t\right)  \cdot\vec{\sigma}$ and $\rho\left(
t\right)  =\left\vert \psi\left(  t\right)  \right\rangle \left\langle
\psi\left(  t\right)  \right\vert =\left[  \mathrm{I}+\mathbf{a}\left(
t\right)  \cdot\mathbf{\sigma}\right]  /2$. Then, we have%
\begin{equation}
\left\langle \mathrm{\dot{H}}\right\rangle =\mathrm{tr}\left[  \rho
\mathrm{\dot{H}}\right]  =\mathbf{a\cdot\dot{m}}\text{.} \label{yo1}%
\end{equation}
Furthermore, using simple quantum mechanics rules, we get
\begin{equation}
\left\langle \mathrm{\dot{H}}\right\rangle =\partial_{t}\left\langle
\mathrm{H}\right\rangle =\partial_{t}\left(  \mathbf{a\cdot m}\right)
=\mathbf{\dot{a}\cdot m+a\cdot\dot{m}}\text{.} \label{yo2}%
\end{equation}
Eqs. (\ref{yo1}) and (\ref{yo2}) seem to be not compatible. However, this is
not the case since $\mathbf{\dot{a}\cdot m}=0$ given that $\mathbf{\dot{a}%
=}2\mathbf{m\times a}$ is orthogonal to $\mathbf{m}$. All we need to do is to
explicitly verify that $\mathbf{\dot{a}=}2\mathbf{m\times a}$. \ Note that,%
\begin{equation}
\left\langle \vec{\sigma}\right\rangle =\mathrm{tr}\left[  \rho\vec{\sigma
}\right]  =\frac{1}{2}\mathrm{tr}\left[  \left(  \mathbf{a\cdot}\vec{\sigma
}\right)  \vec{\sigma}\right]  =\mathbf{a}\text{.} \label{sware}%
\end{equation}
Therefore, using the Schr\"{o}dinger evolution equation $i\hslash\partial
_{t}\left\vert \psi\left(  t\right)  \right\rangle =\mathrm{H}\left(
t\right)  \left\vert \psi\left(  t\right)  \right\rangle $ and setting
$\hslash=1$, the time derivative $\partial_{t}\left\langle \vec{\sigma
}\right\rangle =\mathbf{a}$ of $\left\langle \vec{\sigma}\right\rangle $
in\ Eq. (\ref{sware}) becomes%
\begin{align}
\mathbf{\dot{a}}  &  =\partial_{t}\left\langle \vec{\sigma}\right\rangle
\nonumber\\
&  =\partial_{t}\left(  \left\langle \psi\left\vert \vec{\sigma}\right\vert
\psi\right\rangle \right) \nonumber\\
&  =\left\langle \dot{\psi}\left\vert \vec{\sigma}\right\vert \psi
\right\rangle +\left\langle \psi\left\vert \vec{\sigma}\right\vert \dot{\psi
}\right\rangle \nonumber\\
&  =i\left\langle \psi\left\vert \mathrm{H}\vec{\sigma}\right\vert
\psi\right\rangle -i\left\langle \psi\left\vert \vec{\sigma}\mathrm{H}%
\right\vert \dot{\psi}\right\rangle \nonumber\\
&  =i\left\langle \psi\left\vert \left[  \mathrm{H}\text{, }\vec{\sigma
}\right]  \right\vert \psi\right\rangle \text{,}%
\end{align}
that is, the rate of change of the time-dependent Bloch vector $\mathbf{a}$
reduces to%
\begin{equation}
\mathbf{\dot{a}}=i\left\langle \psi\left\vert \left[  \mathrm{H}\text{, }%
\vec{\sigma}\right]  \right\vert \psi\right\rangle \text{.} \label{a-dot}%
\end{equation}
To further simplify the expression in Eq. (\ref{a-dot}), we note that the
commutator $\left[  \mathrm{H}\text{, }\sigma_{j}\right]  $ can be recast as
\begin{align}
\left[  \mathrm{H}\text{, }\sigma_{j}\right]   &  =\left[  \mathbf{m}\cdot
\vec{\sigma}\text{, }\sigma_{j}\right] \nonumber\\
&  =\left[  m_{i}\sigma_{i}\text{, }\sigma_{j}\right] \nonumber\\
&  =m_{i}\left[  \sigma_{i}\text{, }\sigma_{j}\right] \nonumber\\
&  =m_{i}\left(  2i\epsilon_{ijk}\sigma_{k}\right) \nonumber\\
&  =-2i\epsilon_{ikj}m_{i}\sigma_{k}\text{,}%
\end{align}
that is, for any $1\leq j\leq3$,
\begin{equation}
\left[  \mathrm{H}\text{, }\sigma_{j}\right]  =-2i\epsilon_{ikj}m_{i}%
\sigma_{k}\text{.} \label{b-dot}%
\end{equation}
Therefore, combining Eqs. (\ref{a-dot}) and (\ref{b-dot}), we have%
\begin{align}
\left(  \mathbf{\dot{a}}\right)  _{j}  &  \mathbf{=}i\left(  -2i\epsilon
_{ikj}m_{i}\left\langle \sigma_{k}\right\rangle \right) \nonumber\\
&  =2\epsilon_{ikj}m_{i}a_{k}\nonumber\\
&  =2\left(  \mathbf{m\times a}\right)  _{j}\text{,}%
\end{align}
that is, we arrive at the equality $\mathbf{\dot{a}=}2\mathbf{m\times a}$. As
a conclusive remark, we emphasize that this latter equality implies that
$\partial_{t}\left(  \mathbf{a\cdot m}\right)  =\mathbf{\dot{a}\cdot
m+a\cdot\dot{m}=a\cdot\dot{m}}$ since $\mathbf{\dot{a}\cdot m=}0$ (given that,
by construction, $\mathbf{\dot{a}=}2\mathbf{m\times a}$ is orthogonal to
$\mathbf{m}$).
\end{document}